\documentclass[%
 aip,
 amsmath,amssymb,
 reprint,%
]{revtex4-1}

\usepackage{graphicx}
\usepackage{dcolumn}
\usepackage{bm}

\usepackage[utf8]{inputenc}
\usepackage[T1]{fontenc}
\usepackage{mathptmx}
\usepackage{etoolbox}

\makeatletter
\def\@email#1#2{%
 \endgroup
 \patchcmd{\titleblock@produce}
  {\frontmatter@RRAPformat}
  {\frontmatter@RRAPformat{\produce@RRAP{*#1\href{mailto:#2}{#2}}}\frontmatter@RRAPformat}
  {}{}
}%
\makeatother

\usepackage{xspace}
\usepackage[version=3]{mhchem} 
\usepackage{xcolor}

\newcommand{\red}{\textcolor{black} }

\newcommand*{\wn}{\ensuremath{\text{cm}^{-1}}\xspace}

\newcommand*{\XPi}{\ensuremath{X\,^2\Pi}\xspace}
\newcommand*{\XPiHalf}{\ensuremath{X\,^2\Pi_{1/2}}\xspace}
\newcommand*{\XPiThreeHalf}{\ensuremath{X\,^2\Pi_{3/2}}\xspace}

\newcommand*{\ASigma}{\ensuremath{A\,^2\Sigma^+}\xspace}
\newcommand*{\BSigma}{\ensuremath{B\,^2\Sigma^-}\xspace}

\begin{document}


\title{Elucidating Au-C Bonding via Laser Spectroscopy of Gold Monocarbide}
\author{Rory M. Weldon}
\author{Danielle M. Darling}
\author{Nicole M. Albright}
\author{Kendall L. Rice}
\author{Phaedra L. Salerno}
\author{K. Cooper Stuntz}
\author{Benjamin L. Augenbraun}
\email{bla1@williams.edu}
\affiliation{Department of Chemistry, Williams College, Williamstown, MA 01267, USA}

\date{5 April 2026}

\begin{abstract}
Gold monocarbide (AuC) has been produced and characterized using laser spectroscopy, representing the first reported observation of AuC. We recorded the optical spectrum of gas-phase AuC between 400 nm and 700 nm, assigning excitations from the $\XPiHalf ( (2\sigma)^2 (2\pi)^1 )$ ground state to states arising from the $(2\sigma)^2 (3\sigma^\ast)^1 $ and $(2\sigma)^1 (2\pi)^2 $ configurations. Dispersed-fluorescence spectra are used to study the vibrational and spin-orbit structure of the ground state, branching ratios and radiative lifetimes of the excited states, and the Au--C bond dissociation energy. A molecular orbital diagram is used to rationalize the nature of AuC's low-lying electronic states. The data serve as valuable benchmarks of relativistic theory and are relevant to quantum science and precision measurements with cold molecules.
\end{abstract}

\maketitle

\section{Introduction}
Although bulk gold (Au) is famous for its inertness, many gold compounds are highly efficient catalysts.\cite{skouta2008goldcatalyzed,hashmi2006gold,hashmi2007goldcatalyzed} Both homogeneous and heterogeneous catalysis by gold underlie a staggering number of organic transformations, including industrially important processes. For example, carbon-supported gold catalysts efficiently catalyze the hydrochlorination of acetylene (derived from coal), producing vinyl chloride monomer (VCM) as a precursor to the manufacture of polyvinyl chloride (PVC).\cite{ciriminna2016industrial,johnston2015discovery}
Gold catalysts are also used in the oxidation of cyclohexane to adipic acid (a key precursor in the manufacture of nylons\cite{alshammari2015potential}), the cyclization of alcohols to furans,\cite{liu2005goldcatalyzed} the hydroarylation of alkynes, and many other reactions.\cite{reetz2003goldcatalyzed} Understanding the nature of gold-carbon bonding is critical to clarifying the (often poorly understood) intermediates and mechanisms, and thereby to design promising new catalysts.\cite{hashmi2006gold,hashmi2007goldcatalyzed,benitez2009bonding}

A primary challenge in modeling gold's chemistry is the need for accurate treatment of relativity, which remains a major challenge to theory.\cite{pyykko2012relativistic, pyykko1979relativity, autschbach2012perspective, cheng2014analytic, matthews2020coupledcluster} For instance, proper treatment of relativity is critical to describing gold's anomalously high electronegativity, unusually short bond lengths, stabilized 6s orbitals, and destabilized 5d orbitals.\cite{schwerdtfeger2002relativistic,pyykko2012relativistic,bartlett1998relativistic} These effects are suspected to confer practical benefits, such as gold(I) catalysts' general tolerance of oxygen and/or water and a redox stability that could enable novel reaction schemes.\cite{gorin2007relativistic, dorel2015gold, shapiro2010reactivitydriven, yamamoto2007pelectrophilic} 
Developing frameworks to compute molecular properties including  spin-orbit coupling and other relativistic effects, which can be essential for modeling gold's chemistry, is a major focus of contemporary computational chemistry.\cite{pyykko1979relativity,sherborne2020modular, cheng2011analytic,cheng2014perturbative, zou2011development} High-quality experimental data is necessary to guide and benchmark the development of these relativistic methods and basis sets. 

While realistic homogeneous gold catalysts often comprise large coordination complexes or supported clusters,\cite{skouta2008goldcatalyzed,hashmi2006gold,hashmi2007goldcatalyzed} significant insight can be gained by studying \ce{Au-C} bonding in simplified chemical environments. 
Small molecules can be studied as isolated entities, allowing for a deeper understanding of their fundamental reactivities and generating blueprints for understanding larger analogs.\cite{roithova2009theory,roithova2005gasphase}
Small polyatomic molecules containing \ce{Au-C} bonds have been studied using numerous techniques, including photoelectron spectroscopy (\ce{AuC2-}, \ce{AuCCH-}, and \ce{LAuCCH-} where L = Cl, I, CCH),\cite{visser2013spectroscopic, liu2013probing, wang2017gas, leon2014probing, leon2016probing} microwave spectroscopy (\ce{AuCCH} and \ce{AuCN}),\cite{okabayashi2009detection, okabayashi2013microwave,okabayashi2018fourier} and matrix infrared spectroscopy (\ce{CH3-Au} and \ce{CH3-AuH}).\cite{cho2011infrared, cho2020matrix} These studies have generated important insights relevant to gold catalysis, for instance about how relativistic effects support the favorability of terminal alkynyl-gold bond formation.\cite{liu2013probing}

Experimental study of diatomic AuC---nature's simplest bond between gold and carbon---would represent a valuable baseline to understand gold-carbon bonding in more complex settings. Accurate measurements of low-lying electronic, vibrational, and rotational states in AuC would also provide invaluable benchmarks for relativistic theory methods. Perhaps surprisingly, to the best of our knowledge, experimental observation of diatomic AuC has not been previously reported. 

We also undertook the present work to explore AuC's potential as a probe of fundamental symmetry violation through effects like the electron's electric dipole moment (eEDM).\cite{demille_diatomic_2015} Recent calculations found that AuC, and its heavier congener AuPb, had high intrinsic sensitivity to the eEDM.\cite{stuntz2024optical} The same calculations predicted that AuC had a parity-doubled \XPiHalf ground state that would be robust against common systematic errors, and that the \ce{Au-C} bond length changes very little upon electronic excitation to the \ASigma state.\cite{stuntz2024optical} The minimal change in bond length upon electronic excitation makes AuC an interesting target for optical cycling, the central technique underlying quantum information science and quantum sensing applications with molecules.\cite{fitch2021lasercooled, mccarron2018laser} Confirming the presence of diagonal Franck-Condon factors and a parity-doubled ground state in AuC would be a significant step toward future precision measurements of fundamental symmetry violation using coinage-metal--group-14 dimers.

Here, we report the first experimental study of diatomic AuC. We have produced AuC in a pulsed supersonic molecular beam source and detected the molecules using laser-induced fluorescence spectroscopy. Using a combination of laser excitation and dispersed-fluorescence spectroscopy, we have observed transitions that are assigned to $\ASigma \leftarrow \XPi$ and $\BSigma \leftarrow \XPi$ bands. We have measured the vibrational and spin-orbit structure of the \XPi ground state, vibrational energies and radiative lifetimes of the excited \ASigma and \BSigma states, and the AuC dissociation energy. Franck-Condon factors relevant to optical cycling are also measured using these techniques. We have rationalized this data on the basis of a molecular orbital correlation diagram. Finally, we use this data to benchmark \textit{ab initio} quantum chemical calculations.

\section{Experimental Methods}
Molecular beams of AuC were produced via the reaction of laser-ablated gold vapor with methane (\ce{CH4}). A thick-walled gold tube was ablated by the second harmonic of a pulsed Nd:YAG laser operating at 10~Hz with an ablation energy of approximately 20~mJ. The gold tube was rotated and translated to ensure each ablation pulse hit a fresh surface. A gas mixture of approximately 3\% \ce{CH4} in argon was introduced through a pulsed valve at a backing pressure of 3000~kPa to ensure a low internal temperature. The opening of the pulsed valve was timed to entrain the ablation products before expanding through a nozzle (1~mm diameter, 4~mm long) into a vacuum chamber maintained at a typical running pressure of $5\times 10^{-5}$~Torr.  The supersonic expansion was probed 10~cm downstream from the nozzle by pulsed laser excitation. Three types of experiments were performed: two-dimensional (2D) spectroscopy, dispersed laser-induced fluorescence (DLIF), and radiative lifetime measurements.

Initial survey scans used two-dimensional spectroscopy to search for AuC fluorescence.\cite{reilly2006twodimensional, kokkin_detection_2016} In this technique, a pulsed optical parametric oscillator (OPO) was scanned over the visible range from 400 nm to 700 nm while a 150-nm-wide fluorescence spectral window was monitored. The spectral window was periodically shifted to track the OPO's wavelength throughout the scan. This allows one to record a dispersed-fluorescence spectrum at each laser wavelength, which can be viewed as a 2D image correlating the wavelengths at which molecules are excited against those at which the molecules fluoresce.  Typical excitation pulse energies were in the range 4--8 mJ. Laser-induced fluorescence (LIF) was collected by in-vacuum collection optics consisting of a 2-inch diameter, high-NA condenser lens and a spherical mirror mounted inside a blackened tube. The collected LIF was imaged onto a 0.3~m spectrometer equipped with an intensified charge-coupled device (ICCD). The ICCD gate was triggered at least 100 ns after the laser excitation, which largely eliminated scattered light at the laser wavelength. To further reduce the effects of stray light, background subtraction was performed to remove scattered light from the excitation laser and chemiluminescence from the ablation source. The wavelength axis of the spectrometer was calibrated by recording the emission of both Hg and Ne/Ar lamps.

Radiative lifetimes were measured by tuning the excitation laser to resonance with a particular excitation feature and recording the DLIF spectrum with variable time delay between the pulsed laser and ICCD exposure. The detection gate for the ICCD was typically set to a relatively large value (5 $\mu$s) and the time delay after the excitation was stepped in 100~ns increments. Emission features in the DLIF spectrum were integrated to determine the fluorescence intensity as a function of time. Radiative lifetimes were determined by fitting the fluorescence decay curve to an exponential model.

\section{Results}

\subsection{2D Spectroscopy}

\begin{figure*}[t]
    \centering 
    \includegraphics[width=0.9\textwidth]{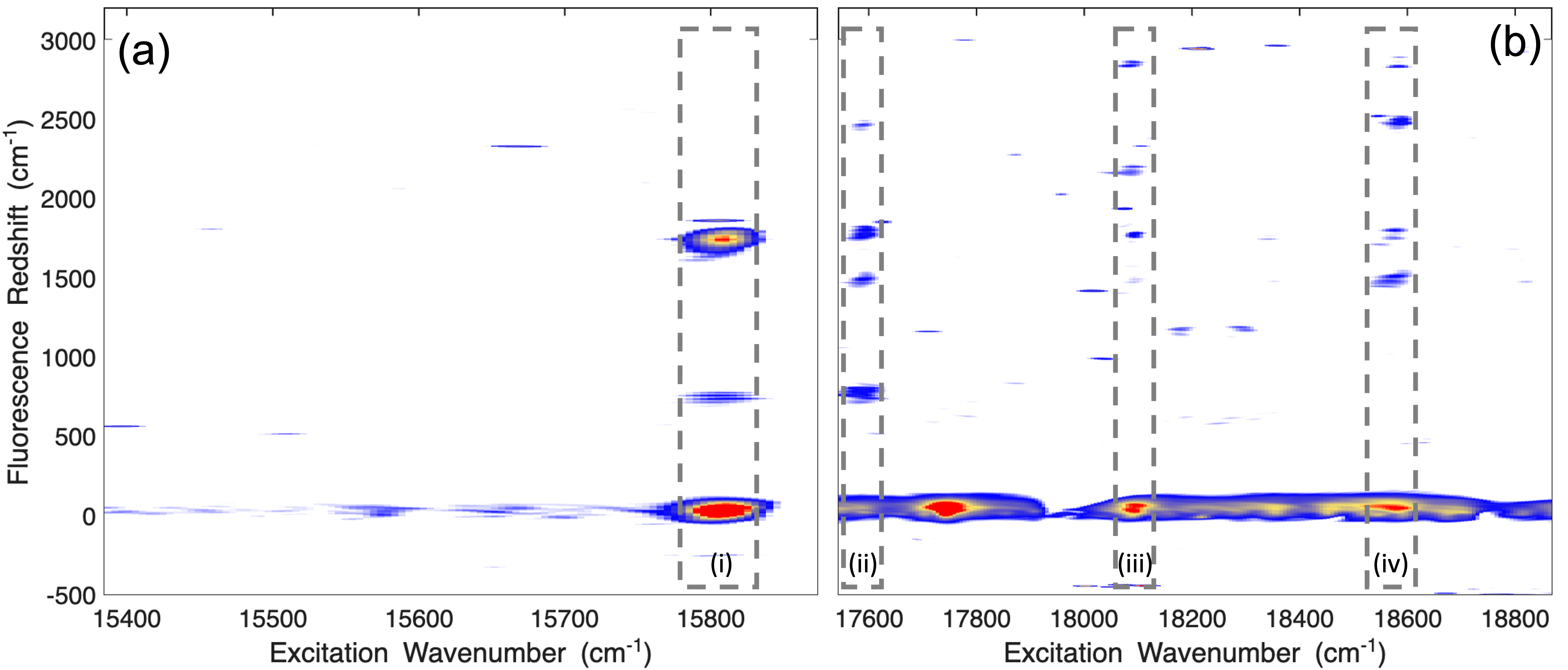}
    \caption{Pulsed-laser 2D spectra recorded near bandheads of the (a) $\ASigma(v'=0) \leftarrow \XPiHalf(v''=0)$ and (b) $\BSigma(v'=0,1,2) \leftarrow \XPiHalf(v''=0)$ transitions. Integration at a fixed excitation wavelength can be used to obtain dispersed fluorescence spectra, while integration at a fixed fluorescence redshift from the laser can be used to obtain an excitation spectrum. Higher-resolution DLIF spectra of regions (i)-(iv) are plotted in Figure \ref{fig:DLIF_AState} and \ref{fig:DLIF_BState}.}
    \label{fig:2DSpectrum}
\end{figure*}

Our initial survey spectroscopy was conducted by monitoring the dispersed fluorescence signal while the excitation laser was scanned from 700 nm to 400 nm. Because no observations of AuC have been reported in the literature, during early scans we sought basic evidence of AuC production via laser ablation. A weak, but repeatable, excitation feature was located near 633 nm, as seen in the 2D spectrum of Figure \ref{fig:2DSpectrum}(a). The signal disappeared when either the Au or the CH$_4$ was removed from the system, implying that the observed molecule contained some combination of Au, C, and/or H. The relatively sparse vibrational progression observed in the 2D spectrum made us suspect this originated from a diatomic molecule, and the only reasonable candidate with a vibrational interval of $\sim$700 \wn is AuC. The other redshifted fluorescence feature in Figure \ref{fig:2DSpectrum}(a) does not occur at a simple multiple of the fundamental vibrational frequency and thus was suspected to come from decay to a low-lying metastable state. Since AuC is predicted to have a $^2\Pi_{1/2}$ ground state, this matches expectations for a spin-orbit-split $^2\Pi_{3/2}$ state a few thousand \wn higher in energy.\cite{stuntz2024optical} 

Having established a strong likelihood that we had produced AuC, we continued the 2D spectroscopic survey toward higher excitation energies. We successfully located vibrational excitations associated with the electronic state that gave rise to the 633 nm feature described above. This ``red'' series of transitions had excitations near 633 nm, 605 nm, 581 nm, and 559 nm---consistent with a series of vibrational levels separated by $\sim$700~\wn. At yet higher energies, a second electronically-excited state was located. This ``blue'' series had vibronic transitions near 568 nm, 552 nm, 537 nm, 524 nm, and 511 nm; these seemed to comprise a vibrational progression with excited-state levels separated by about 500~\wn. A representative portion of this data, which ultimately led to the identification of the \BSigma state, is shown in Figure \ref{fig:2DSpectrum}(b). Ultimately, these progressions could be assigned to transitions among the $X\,^2\Pi(v''=0-7)$, $\ASigma(v'=0-3)$, and $\BSigma(v'=0-4)$ states. A complete listing of the observed transition wavenumbers is provided in Table S1 of the Supplementary Material.

\subsection{Dispersed Laser-Induced Fluorescence}
\begin{figure*}[t]
    \centering 
    \includegraphics[width=0.89\textwidth]{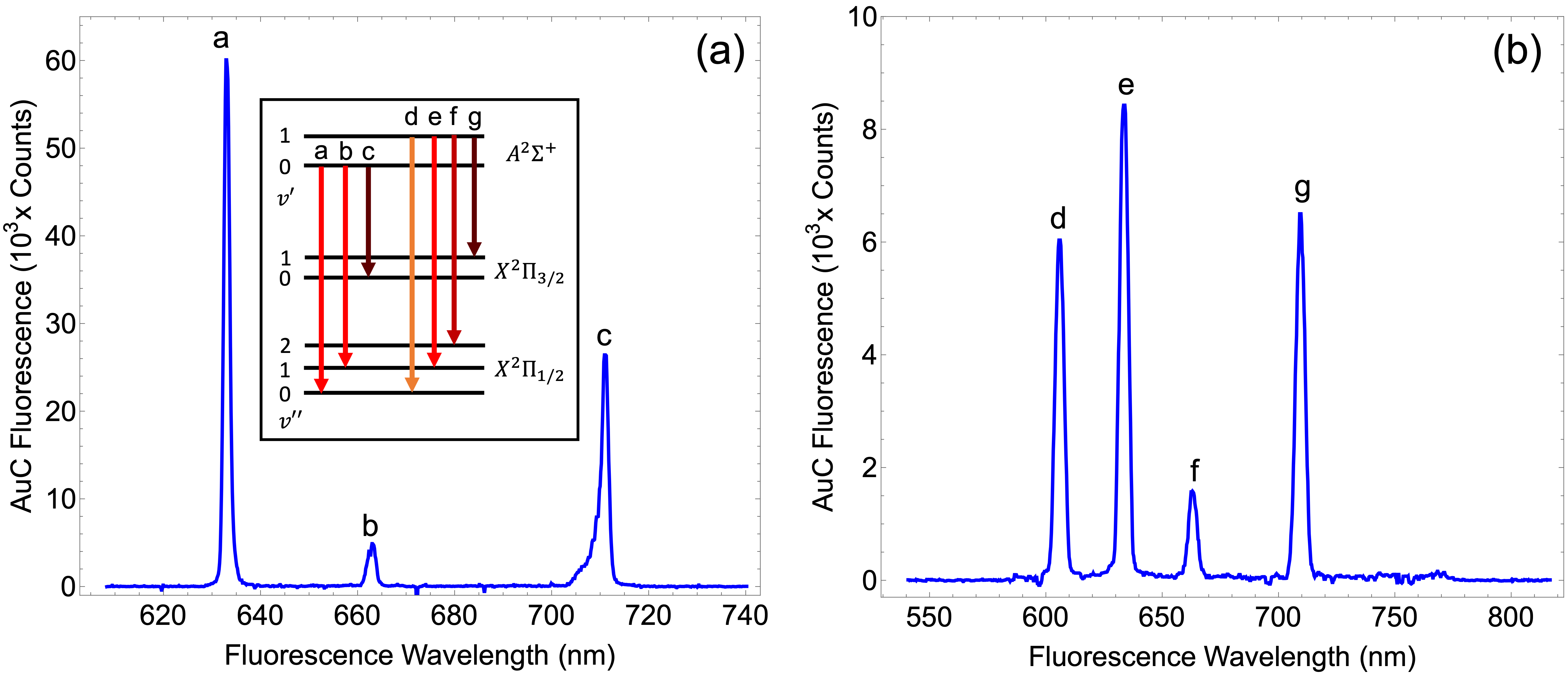}
    \caption{Dispersed fluorescence spectra resulting from laser excitation near the bandheads of the (a) $\ASigma(v'=0) \leftarrow \XPiHalf(v''=0)$ and (b) $\ASigma(v'=1) \leftarrow \XPiHalf(v''=0)$ transitions. The inset shows an energy level diagram with vibrational and spin-orbit assignments for each observed transition.}
    \label{fig:DLIF_AState}
\end{figure*}

Higher-resolution DLIF spectra were recorded by fixing the OPO's excitation wavelength to each bandhead observed via 2D spectroscopy and using a higher resolution grating (600 lines/mm) in the spectrometer. A typical background-subtracted DLIF spectrum consisted of accumulating the fluorescence signal from 7500 ablation pulses. DLIF spectra recorded for excitation features in the ``red'' vibrational progression are shown in Figure \ref{fig:DLIF_AState}. As can be seen, these excited states display strong diagonal fluorescence near 633 nm with a few weaker features spaced by approximately 700~\wn, indicative of the ground-state vibrational spacing. Within the $\ASigma \rightarrow \XPiHalf$ subband, the Franck-Condon factors are evidently very diagonal, making it difficult to observe decays with $\Delta v>1$.  A prominent feature approximately 1750~\wn to the red of the diagonal fluorescence represents decay to the \XPiThreeHalf state, which is separated from the \XPiHalf ground state due to spin-orbit coupling. Interestingly, Figure \ref{fig:DLIF_AState}(a) shows that this feature is strongly degraded to shorter wavelengths in the case of excitation to \ASigma$(v'=0)$, which we hypothesize is due to the underlying rotational contour. Future high-resolution studies may help explain this observation.   We assign the excitation features of this ``red'' progression to the $\ASigma - \XPi$ transition, since that transition was predicted to feature highly diagonal Franck-Condon factors.\cite{stuntz2024optical} We have performed additional excited-state computations (described below) to further support this assignment.

\begin{figure*}[t]
    \centering 
    \includegraphics[width=0.89\textwidth]{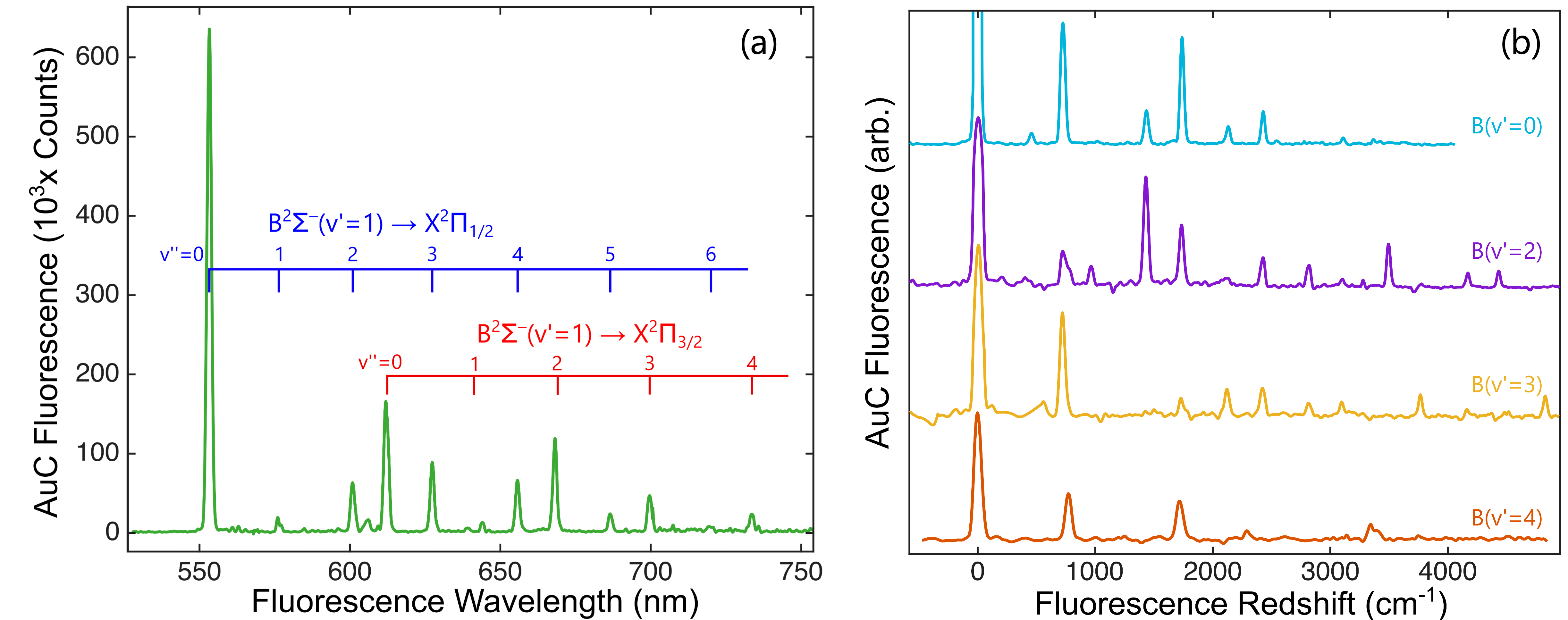}
    \caption{(a) Dispersed fluorescence spectra resulting from laser excitation near the bandhead of the (a) $\BSigma(v'=1) \leftarrow \XPiHalf(v''=0)$ transition. The vibrational and spin-orbit quantum number assignments are shown above the data. (b) Comparison of dispersed fluorescence spectra recorded following excitation of the $\BSigma (v=0,2,3,4)$ states. The diagonal peak is off-scale so as to emphasize signal near the baseline.}
    \label{fig:DLIF_BState}
\end{figure*}

DLIF spectra recorded for excitation features in the ``blue'' vibrational progression are shown in Figure \ref{fig:DLIF_BState}. In contrast to the DLIF spectra associated with the \ASigma state, the excited states in the ``blue'' progression show strongly off-diagonal fluorescence. This implies a significant change in bond length between the ground and excited states of these transitions. This is advantageous to our data analysis in that we observe fluorescence features up to $v''=7$, providing high quality determination of the ground-state vibrational and spin-orbit structure. Based on the estimated vibrational and spin-orbit energies obtained from analysis of the $\ASigma \rightarrow \XPi$ spectra, it was straightforward to assign the vibronic quantum numbers for all observed fluorescence features. We assign this progression to the \BSigma electronically excited state. As will be discussed below, this is supported by expectations from a qualitative molecular orbital diagram and is supported by \textit{ab initio} calculations, which predict a significant change in bond length and vibrational frequency between \XPiHalf and \BSigma.

A global fit of 60 observed vibronic bands was performed to determine term energies ($T_e$), vibrational frequencies ($\omega_e$), and anharmonic contributions ($\omega_e x_e$). The vibrational energy levels were modeled using the standard expression
\begin{equation}
    T_v = T_e + \omega_e (v+1/2) - \omega_e x_e (v+1/2)^2,
\end{equation}
with all parameters expressed in \wn. The fitted values of $T_e$, $\omega_e$, and $\omega_e x_e$ for the \XPiHalf, \XPiThreeHalf, \ASigma, and \BSigma states are reported in Table \ref{tab:VibrationalFit}. The optimized parameters can be used to predict the differences between observed and calculated transition wavenumbers listed in Table S1 of the Supplementary Material. The overall standard deviation of the fit was 4 \wn, commensurate with the spectrometer's resolution.

\begin{table}[t]
\begin{center}
    \caption{Vibrational parameters for AuC determined from the global band fit. Values in parentheses represent $1\sigma$ uncertainties. All values are reported in units of \wn.}
		\label{tab:VibrationalFit}
    \begin{ruledtabular}
		\begin{tabular}{ccccc}
 State   & $T_e$ & $\omega_e$ & $\omega_e x_e$  \\ \hline
 \XPiHalf & 0 & 726.6(1.4) & 4.43(0.20)  \\
 \XPiThreeHalf & 1746.7(3.5) & 697.5(3.5) & 4.20(0.64) \\
 \ASigma & 15805.3(4.8) & 718.1(5.8) & 7.2(1.5) \\
 \BSigma & 17671.3(3.2) & 516.2(2.3) & 6.23(0.45)  \\
		\end{tabular}
    \end{ruledtabular}
\end{center}
\end{table}

\subsection{Vibrational Branching Ratios}
Vibrational branching ratios can be determined from our DLIF spectra by integrating over a small range of wavelengths centered on each decay feature and comparing to the total integrated fluorescence for all decays. Branching ratios are of particular interest for the identification of optical cycling transitions, in which an excited electronic state displays strongly suppressed ``off-diagonal'' decays that change the vibrational quantum number. Such transitions are one of the key enabling technologies behind quantum state control of molecules.\cite{fitch2021lasercooled, mccarron2018laser} Figures \ref{fig:DLIF_AState} and \ref{fig:DLIF_BState} reveal that the $\ASigma \rightarrow \XPi$ transition enjoys this property, while the $\BSigma \rightarrow \XPi$ band does not. For that reason, we focus here on the decays from the \ASigma state. For the $\ASigma \rightarrow \XPiHalf$ subband, the data of Figure~\ref{fig:DLIF_AState} yields branching ratios of approximately 93(1)\% and 7(1)\% to $\XPiHalf(v''=0)$ and $\XPiHalf(v''=1)$, respectively. This compares quite favorably to recent theoretical predictions.\cite{stuntz2024optical} Our data also provides the relative intensities of decays on the $\ASigma \rightarrow X\,^2\Pi_{1/2,3/2}$ subbands. The $\ASigma \rightarrow \XPiThreeHalf$ decay has a branching ratio of 33(3)\%, while the $\ASigma \rightarrow \XPiHalf$ subband has a branching ratio of 67(2)\%. Combining calculated H\"{o}nl-London factors and the frequency dependence of the transition rate, one predicts branching fractions of 0.41 and 0.59 for decays to \XPiThreeHalf and \XPiHalf, respectively, which is in reasonable agreement with the measured values. For the $\ASigma \rightarrow \XPi$ bands, transitions with $\Delta v\geq2$ were not observed above the 0.1\% level. This is very promising for the development of quasi-closed optical cycling transitions needed for quantum-state control,\cite{fitch2021lasercooled, mccarron2018laser, hutzler2020polyatomic} since it appears that application of just three laser wavelengths (to address the $\XPiHalf(v''=0,1)$ and $\XPiThreeHalf(v''=0)$ states) may be sufficient to scatter $\sim$100 photons per AuC molecule.

\subsection{Radiative Lifetimes}

\begin{figure}[tb]
    \centering 
    \includegraphics[width=1\columnwidth]{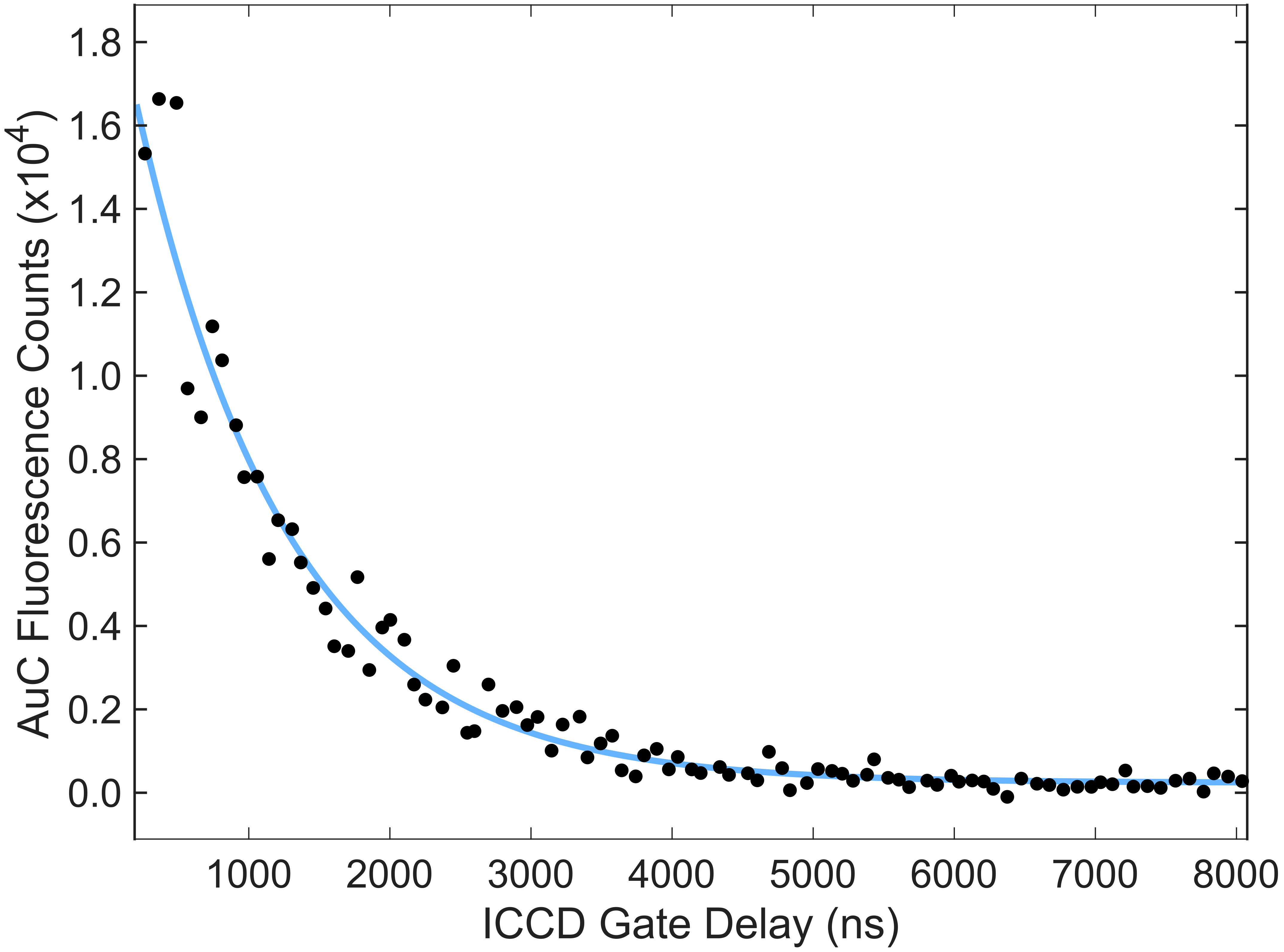}
    \caption{Fluorescence decay curve following laser excitation near the bandhead of the $\BSigma(v'=0) \leftarrow \XPiHalf(v''=0)$ transition at 568.2 nm. The blue line represents an exponential fit to the data.}
    \label{fig:lifetimes}
\end{figure}

The radiative lifetimes of the observed AuC excited states were measured by fixing the OPO's wavelength to the bandhead of a particular vibronic transition and recording the DLIF spectrum at variable time delays after the pulsed-laser excitation. The ICCD gate (5 $\mu$s wide) was delayed after the excitation laser pulse in steps of 50 ns. The resulting spectra were integrated within $\pm$10~nm about the resonant fluorescence and fit to an exponential decay model to determine the excited-state lifetimes. A representative fluorescence decay curve recorded following excitation of the $\BSigma(v'=0)$ level is shown in Figure \ref{fig:lifetimes}. This curve comes from the accumulation of 50 images at each time step. The fitted radiative lifetimes are determined to be 1340(70)~ns for the $\ASigma(v'=0)$ state, 1080(70)~ns for the $\BSigma(v'=0)$ states. For the purposes of molecular laser cooling, one typically seeks lifetimes $\tau \lesssim 100$~ns, so the long radiative lifetimes of the AuC excited states may prohibit direct laser cooling. Nonetheless, they are still sufficiently short to allow facile quantum state preparation.\cite{mccarron2018laser, fitch2021lasercooled}

\section{Discussion}

\subsection{Rationalization of the Electronic States}

\begin{figure}[tb]
    \centering 
    \includegraphics[width=0.9\columnwidth]{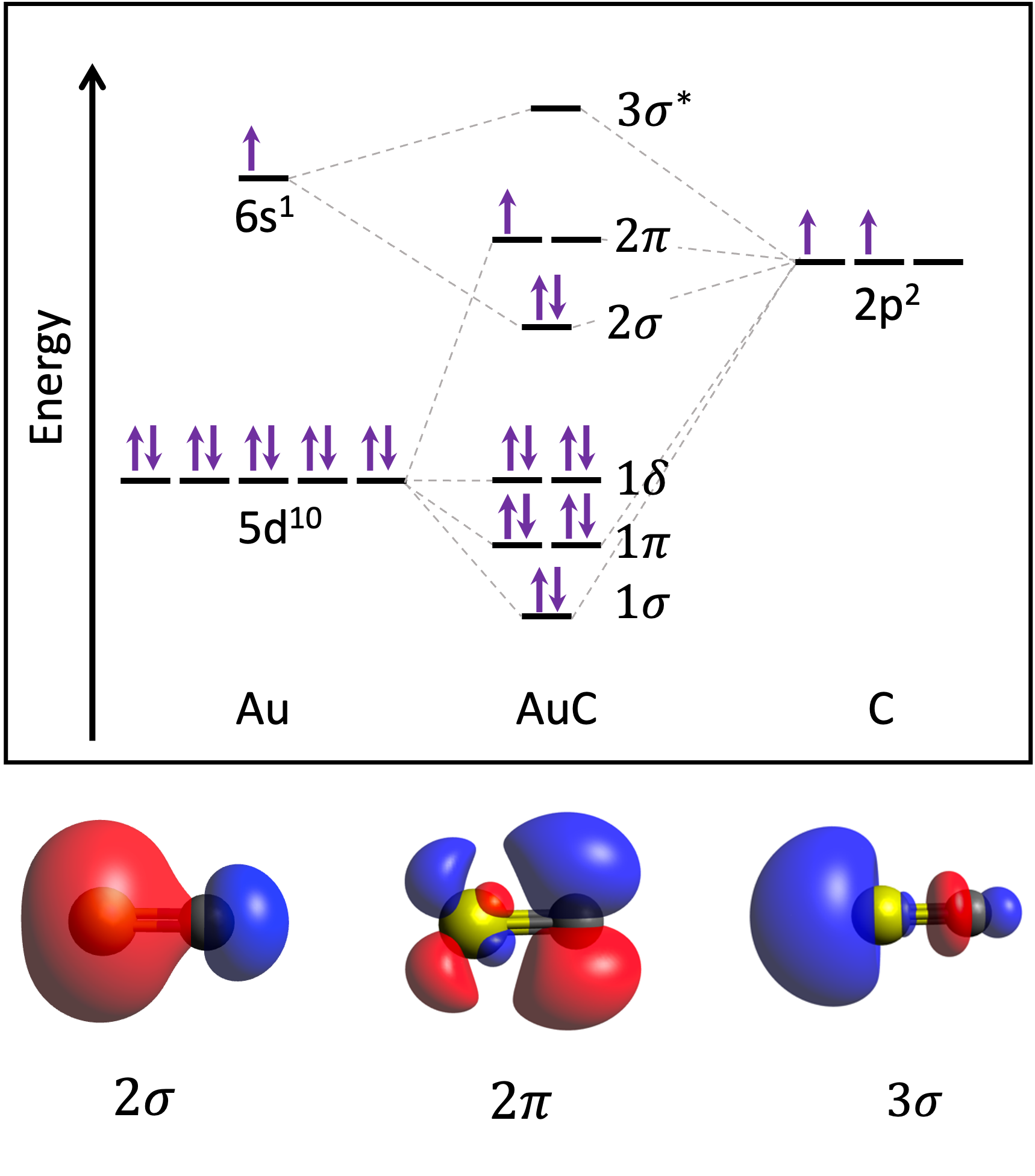}
    \caption{Top: Molecular orbital correlation diagram to rationalize the low-lying electronic states in AuC. Bottom: renderings of the HOMO-1 ($2\sigma$), HOMO ($2\pi$), and LUMO ($3\sigma^\ast$) orbitals. The molecules are oriented with Au on the left and C on the right.}
    \label{fig:OrbitalDiagram}
\end{figure}

Interpretation of the electronic structure of AuC can be guided by a qualitative molecular orbital correlation diagram shown in Figure~\ref{fig:OrbitalDiagram}.\cite{kokkin2015au,obrien2008intracavity} The $1\sigma$, $1\pi$, and $1\delta$ molecular orbitals are primarily Au 5d in character, with a small contribution from C 2p to $1\sigma$ and $1\pi$. The bottom of Figure \ref{fig:OrbitalDiagram} shows renderings of the frontier molecular orbitals. The $2\sigma$ molecular orbital is a bonding combination of Au 6s and C 2p. The $2\pi$ orbital is a predominantly C-centered orbital that is slightly antibonding with some Au 5d$\pi$ character. Finally, the $3\sigma^\ast$ molecular orbital is a primarily Au-centered 6s/6p back-polarized orbital that has an antibonding contribution from C 2p. This pattern leads to a ground electronic configuration of $(1\sigma)^2(1\pi)^4(1\delta)^4(2\sigma)^2(2\pi)^1$, which gives rise to a single regular ${}^2\Pi$ state (i.e., ${}^2\Pi_{1/2}$ below ${}^2\Pi_{3/2}$).

The measured spin-orbit splitting of the \XPi state can be used to estimate the contribution of $5d \pi_\mathrm{Au}$ and $2p \pi_\mathrm{C}$ atomic orbitals to the $2\pi$ molecular orbital. Two assumptions are made for this estimate: that only the Au $5d$ and C $2p$ atomic orbitals contribute to this molecular orbital, and that the AuC \XPi state is appropriately modeled by a ${}\cdots (2\sigma)^2 (2\pi)^1$ single configurational wavefunction.\cite{dabell2001electronica, lefebvre-brion2004spectra} In this approximation, we write the $2\pi$ molecular orbital as 
\begin{equation}
    \lvert 2\pi \rangle = \alpha \lvert 5d\pi_\mathrm{Au} \rangle + \beta \lvert 2p \pi_\mathrm{C} \rangle.
\end{equation}
The molecular spin-orbit parameter is then expressed as~\cite{lefebvre-brion2004spectra}
\begin{equation}
    A = a_\pi = \langle 2\pi \rvert \hat{a} \lvert 2\pi \rangle = \lvert \alpha \rvert^2 \zeta_{5d\mathrm{(Au)}} + \lvert \beta \rvert^2 \zeta_{2p\mathrm{(C)}}.
\end{equation}
Estimates of the atomic spin-orbit parameters are $\zeta_{5d\mathrm{(Au)}} = 5100$~\wn and $\zeta_{2p\mathrm{(C)}} = 32$~\wn.\cite{houdart1973emission,dabell2001electronica} Combining these values with the measured $A = 1750$~\wn shows that the contributions of the Au and C orbitals are approximately 34\% and 66\%, respectively. This estimate assumes negligible overlap between the Au and C orbitals; including some amount of overlap tends to increase the magnitude of $\beta$ without significantly altering the value of $\alpha$. The large value of $\zeta_{5d\mathrm{(Au)}}$ means that $\alpha$ is determined rather restrictively. By contrast, the small value of $\zeta_{2p\mathrm{(C)}}$ means that $\beta$ could vary without significantly changing the weighted average. 
The 33\% Au contribution to the $2\pi$ orbital is in fair agreement with \textit{ab initio} calculations predicting 40\% Au contribution to the highest occupied molecular orbital (HOMO).\cite{stuntz2024optical}

We now turn to the electronically excited states. The lowest energy excitations from the ground-state electron configuration would likely be of the form  $2\pi \rightarrow 3\sigma^\ast$ or $2\sigma \rightarrow 2\pi$. These promotions generate five electronically excited states:
\begin{align*}
    &(1\sigma)^2(1\pi)^4(1\delta)^4(2\sigma)^2(2\pi)^0(3\sigma^\ast)^1 \Rightarrow A\,^2\Sigma^+\\ 
    &(1\sigma)^2(1\pi)^4(1\delta)^4(2\sigma)^1(2\pi)^2 \Rightarrow a\,^4\Sigma^-, B\,^2\Sigma^-, C\,^2\Delta, D\,^2\Sigma^+
\end{align*}
Qualitative inspection of the molecular orbital renderings (at the bottom of Figure \ref{fig:OrbitalDiagram}) reveals that the $2\pi$ and $3\sigma^\ast$ MOs exhibit similar degrees of antibonding character, so the \ASigma state is expected to have a vibrational frequency similar to that of the ground state. By contrast, the $2\sigma$ MO is strongly bonding, so the $2\sigma \rightarrow 2\pi$ promotion will generate excited states with significantly lower vibrational frequency. On this basis, we assign the vibrational progression beginning at 633 nm as \ASigma and the progression beginning at 568 nm as \BSigma. \textit{Ab initio} estimates\cite{stuntz2024optical, li2011electronic} suggest that the $a\,^4\Sigma^-$ state is $<9,000$~\wn above the ground state, outside our observation window. While we searched for the $C\,^2\Delta$ and $D\,^2\Sigma^+$ states at wavelengths down to 410 nm, we were unable to detect evidence of AuC fluorescence in this region.

The fitted value of the \ASigma vibrational frequency is  \red{718(6)~\wn}, quite similar to the ground-state vibrational frequency of \red{727(1)~\wn}. The similarity between these values suggests that the degree of (anti)bonding character does not change substantially upon electronic excitation $2\pi \rightarrow 3\sigma^\ast$. Well-matched vibrational frequencies were predicted by the previous theoretical study of AuC.\cite{stuntz2024optical} Moreover, the measured values agree quite favorable with the theoretical predictions of 732 \wn and 726 \wn for the \XPiHalf and \ASigma states, respectively.\cite{stuntz2024optical} By contrast, the fitted value of the \BSigma vibrational frequency is significantly reduced, \red{$\omega_e(\BSigma) = 516(2)$} \wn. This can also be rationalized on the basis of the MOs and qualitative energy level diagram of Figure \ref{fig:OrbitalDiagram} because the \BSigma state arises from excitation of an electron from the strongly bonding $2\sigma$ orbital to the antibonding $2\pi$ orbital. 

The radiative lifetime of an excited vibronic level, $\tau_{iv'}$, and its branching ratios, $b_{iv',fv''}$, are related to the Einstein A spontaneous emission coefficient between initial ($i$) and final ($f$) vibronic levels via 
\begin{equation}
    A_{iv',fv''} = b_{iv',fv''} \, \tau_{iv'}^{-1}.
\end{equation}
The Einstein A coefficient can also be related to the transition dipole moment (TDM) as
\begin{equation}
    A_{iv',fv''} = 3.137\times10^{-7} \lvert \mu_{iv',fv''} \rvert^2 \, \nu_{iv',fv''}^3,
\end{equation}
where the transition dipole moment (TDM), $\mu_{iv',fv''}$ is in Debye (D) and the transition frequency, $\nu_{iv',fv''}$ is in wavenumbers (\wn). These two relationships allow us to combine the experimentally measured transition frequencies, radiative lifetimes, and branching ratios to deduce the TDMs $\mu_{A\leftarrow X} = 0.62(2)$~D and $\mu_{B\leftarrow X} = 0.52(3)$~D. These values are in good agreement with the computations described below, which predict both electronic transitions have transition dipole moments around 0.6~D.

To support our assignments of the excited electronic configurations, we have run a set of time-dependent density functional theory (TD-DFT) calculations in the ORCA computational package using unrestricted Kohn-Sham (UKS) TD-DFT with the B3LYP functional.\cite{neese_software_2025} This allowed computation of excitation energies and geometry optimization using analytic Hessians. Vibrational frequencies were determined via numerical differentiation. Relativistic effects were included using ORCA's X2C Hamiltonian and a segmented contracted triple-$\zeta$ basis set with one set of polarization functions on the heavy atoms.\cite{pollak_segmented_2017, franzke_error-consistent_2019} These calculations predict a $\pi \rightarrow \sigma$ excitation at approximately 16,400~\wn and a $\sigma \rightarrow \pi$ excitation near 17,800~\wn. Vibrational frequencies associated with these two excited states are calculated to be 713 \wn for the first excited state and 529 \wn for the second excited state. Both of these frequencies are in good agreement with the measured values of \red{$\omega_e(\ASigma) = 718(6)$} \wn and \red{$\omega_e(\BSigma) = 516(2)$} \wn, which lends further support to our assignments. The transition dipole moments are computed as $\mu_{A\leftarrow X} = 0.66$ D and $\mu_{B\leftarrow X} = 0.63$ D, which is consistent with the radiative lifetime measurements reported above.

\subsection{Gold-Carbon Bond Strength}

Assuming that a Morse potential appropriately describes the AuC ground state, we can use the measured values of $\omega_e$ and $\omega_ex_e$ to calculate the dissociation energy using the relation $D_e = \omega_e^2/4\omega_ex_e$. The standard correlation rules dictate that the AuC \XPiHalf ground state dissociates to the $\mathrm{Au}(^2\mathrm{S}_{1/2})+\mathrm{C}(^3\mathrm{P}_0)$ limit while the \XPiThreeHalf  and \BSigma states dissociate to $\mathrm{Au}(^2\mathrm{S}_{1/2})+\mathrm{C}(^3\mathrm{P}_1)$. Thus, by adding the term energy ($T_e$) and subtracting the spin-orbit energy of the carbon atom from the computed value of $D_e$, we may estimate the molecular dissociation energy. 
We find $D_e(\XPiHalf)=29\,794 (60)$ \wn, $D_e(\XPiThreeHalf)=30\,689 (150)$ \wn, and $D_e(\BSigma)=28\,347 (125)$ \wn. Averaging these values, we obtain a dissociation energy of $3.67(1)$ eV. The values in parentheses represent the $1\sigma$ uncertainty due only to statistical uncertainty. 
Because the experimental estimate of $D_e$ is based on a Morse potential extrapolation from fairly low-lying vibrational levels ($v \leq 7$), it is almost certainly an overestimate of the dissociation energy. Nonetheless, it compares very favorably with the CCSD(T) predicted value of 3.41 eV.\cite{li2011electronic}

\subsection{Comparison to \textit{Ab Initio} Theory} \label{sec:AbInitio}
Over the years, several theoretical treatments of AuC have been published in the literature. Our measurements represent the first opportunity to test these predictions against experimental data, as is done in Table \ref{tab:TheoryComparison}. Accurate treatment of both electron correlation and relativistic effects is essential to describing the electronic structure of AuC. This is due to the fact that there is energetic competition between ${}^2\Pi$ and ${}^4\Sigma$ terms to be the ground state.\cite{li2011electronic, wang2007theoretical} Only upon high-level treatment of both electron correlation and relativistic effects is the ${}^2\Pi$ term found to be the ground state, falling approximately 1.1 eV lower in energy than the ${}^4\Sigma^-$ term.\cite{li2011electronic} Our observation of a ground state that can be assigned as \XPi validate this prediction. Li et al. (Ref.~\onlinecite{li2011electronic}) have also predicted that the \ce{Au-C} bond is quite strong ($D_e \approx 3.4$~eV), with a high vibrational frequency ($\omega_e \approx 780$~\wn) and a relativistically contracted bond length ($r_e \approx 1.8$~\AA). These predictions are also in good agreement with our measurements. Future studies of the rotational structure of AuC are now strongly motivated so that the bond length can be compared to theoretical predictions.

With regards to excited states, previously published EOMEA-CCSD calculations using the exact two-component theory with atomic mean-field integrals (X2CAMF) to treat relativistic effects\cite{dyall_interfacing_2001, Liu2018} were used to predict the energies of the \XPiThreeHalf and \ASigma states.\cite{stuntz2024optical} These states were computed to be at 2299~\wn and 14475~\wn above \XPiHalf, respectively. The 0.16~eV error in the EOMEA-CCSD prediction for the \ASigma energy is substantial, but not beyond the expected accuracy of $\sim$0.2~eV for EOM-CCSD excitation energies. The experimental value of the spin-orbit splitting (\red{1747(4)~\wn}) deviates notably from the EOM-CCSD prediction (2299~\wn). 

To investigate this discrepancy, we performed new CCSD(T) calculations. Since the \XPiHalf and \XPiThreeHalf states are the lowest electronic states with their respective $\Omega$ values, they can be optimized directly in Kramers unrestricted Hartree-Fock (UHF) calculations. We have therefore performed UHF-based CCSD augmented with noniterative triples [CCSD(T)] calculations\cite{raghavachari_fifth-order_1989} for these two states. The UHF-CCSD(T) calculations directly optimize the molecular spinors for the targeted states and are likely to provide more accurate predictions of energies and properties than the previously reported EOMEA-CCSD calculations. The calculations of potential energy surfaces have used Dyall’s correlation-consistent triple-$\zeta$ (cc-pVTZ-SO) basis sets for Au recontracted for the X2CAMF scheme.\cite{dyall_relativistic_2004} These X2CAMF basis sets feature separate contraction coefficients for each spin-orbit component~\cite{zhang_new_2024} and thus can account for spin-orbit coupling effects in heavy elements. For C, correlation-consistent basis sets recontracted for the spin-free X2C theory in its one-electron variant (the SFX2C-1e scheme)~\cite{jr_gaussian_1989, dyall_interfacing_2001, liu_exact_2009} were used since spin-orbit coupling is relatively weak.

Our CCSD(T) calculations predict an energy splitting of 1690~\wn between the minima of the \XPiHalf and \XPiThreeHalf potentials, which agrees much better with the measured value of \red{1747(4)~\wn}. We hypothesize that the improved agreement is because the CCSD(T) calculations more accurately capture wavefunction relaxation than the EOM-CCSD calculations. These results also implies that the relaxation differs between the $\Pi_{1/2}$ and $\Pi_{3/2}$ components, perhaps due to the fact that the $\Omega=1/2$ component can mix with excited $\Sigma$ states. Our measurements thus serve as an important benchmark of both relativistic and electron correlation effects in heavy molecules. The UHF-CCSD(T) calculations also accurately capture the spin-orbit dependence of the vibrational frequency, which is calculated to be slightly reduced in \XPiThreeHalf ($\omega_e = 687$~\wn) relative to \XPiHalf ($\omega_e = 717$~\wn). These values agree quite favorably with experimental values of \red{$698(4)$~\wn} and \red{$727(2)$~\wn}, respectively. 

\begin{table}[t]
\begin{center}
    \caption{Comparison of experimental and theoretical properties for the \XPi state of AuC.}
		\label{tab:TheoryComparison}
    \begin{ruledtabular}
		\begin{tabular}{ccccc}
 Method   & $\omega_e$ (\wn) & $A$ (\wn) & $D_e$ (eV) & $r_e$ (\AA)  \\ \hline
 Expt.\footnote{Present work. $D_e$ computed using Morse potential model.} & 726.6(1.4) & 1746.7(3.5) & 3.67(1) &   \\
 UHF-CCSD(T)$^\mathrm{a}$ & 717 & 1690 &  2.95 & 1.831 \\
EOM-CCSD\footnote{From Ref. \onlinecite{stuntz2024optical}. Equations-of-motion electron-attachment CCSD theory. The X2CAMF scheme was used to treat relativistic effects.} & 732 & 2299 &  3.77 & 1.815  \\
CCSD(T)/RPP\footnote{From Ref. \onlinecite{li2011electronic}. CCSD(T) with relativistic pseudopotentials.} & 781.7 &   & 3.41 & 1.805 \\
CCSD(T)/NRPP\footnote{From Ref. \onlinecite{li2011electronic}. CCSD(T) with nonrelativistic pseudopotentials.} & 430.9 &   & 1.42 & 2.095 \\
DFT\footnote{From Ref. \onlinecite{wang2007theoretical}. Density functional theory with the B3LYP functional.} & 654 &   & 2.65 & 1.866
		\end{tabular}
        \vspace{-1em}
    \end{ruledtabular}
\end{center}
\end{table}

\section{Conclusion}
We have detected and characterized gold monocarbide (AuC) using laser spectroscopy of gas-phase molecules. We have characterized the fine and vibrational structure of the lowest electronic transitions in AuC. We find that AuC has a \XPiHalf ground state, a relatively high vibrational frequency, and a significant dissociation energy. The nature of the low-lying electronic states has been rationalized using a simple molecular orbital diagram. This work provides numerous measurements that can be used to assess the accuracy of computational methods, especially for capturing the importance of relativistic effects. Our measurements lay the groundwork for future high-resolution studies to measure hyperfine structure and electric dipole moments in the low-lying states of AuC.

This work paves the way toward a detailed understanding of how to design molecules to search for fundamental CP violation. Our work has established the presence of a parity-doubled absolute ground state that would enable robust rejection of systematic errors in searches for the electron's electric dipole moment.\cite{acme_collaboration_improved_2018, demille_diatomic_2015} The $\ASigma \rightarrow \XPi$ transition appears promising for the possibility of optical cycling. Although the $\mu$s-scale excited-state lifetimes may limit the possibility of direct laser cooling, the ability to establish an optical cycling transition still enables quantum state preparation and high-efficiency readout.\cite{hutzler2020polyatomic, mccarron2018laser, fitch2021lasercooled} These results also establish a baseline to understand the behavior of heavier congeners that are predicted to have high intrinsic sensitivity to the electron's electric dipole moment.\cite{stuntz2024optical} AuPb is a particularly exciting target for future precision measurements because it consists of two individually laser-coolable atoms,\cite{dzuba_time_2021} which opens the possibility of assembling ultracold molecules from these atoms.\cite{fleig_theoretical_2021, Klos2022} This makes the realization of dense samples of ultracold AuPb a tantalizing prospect. Our lab is currently investigating the structure of AuPb to understand its viability for future searches for the electron's electric dipole moment.

\begin{acknowledgments}
We are grateful to Tim Steimle (Arizona State University) for valuable conversations about this work. We thank Lan Cheng (The Johns Hopkins University) for guidance on the \textit{ab initio} calculations. Finally, we thank Jason Mativi and Dave Williams for technical support. This work was supported by the National Science Foundation under Grant No. PHY-2513425. We also acknowledge support by the ACS Petroleum Research Fund under Undergraduate New Investigator Grant \#67451-UNI6. 
\end{acknowledgments}

\section*{Data Availability Statement}
The data that supports the findings of this study are available within the supplementary material.

\bibliography{AuC-lowres-library}

\end{document}